\documentclass[aps,twocolumn,prl,superscriptaddress,amsmath,amssymb,amsfonts]{revtex4-1} 
\usepackage{amssymb,amsmath}
\usepackage{graphicx}
\usepackage{subfigure}
\usepackage[english]{babel}
\usepackage{float}
\usepackage{color}
\usepackage{booktabs}
\usepackage{blindtext}
\usepackage{comment}
\usepackage[symbol]{footmisc}

\usepackage{tikz,pgf}
\usepackage[normalem]{ulem}
\usepackage{cancel}
\usepackage{hyperref}
\hypersetup{
    colorlinks=true,
    linkcolor=blue,
    citecolor=blue,
    filecolor=cyan,      
    urlcolor=cyan,
}
\usepackage{notes2bib}
\bibnotesetup{
note-name = ,
use-sort-key = false
}
\begin{document}
\title{Dynamical beats of short pulses in waveguide QED}
\author{Dianqiang Su} 
\affiliation{State Key Laboratory of Quantum Optics and Quantum Optics Devices, Institute of Laser Spectroscopy,
Shanxi University, Taiyuan 030006, People's Republic of China.}
\affiliation{Collaborative Innovation Center of Extreme Optics, Shanxi University, Taiyuan 030006,
People's Republic of China.}
\author{Yuan Jiang} 
\affiliation{State Key Laboratory of Quantum Optics and Quantum Optics Devices, Institute of Laser Spectroscopy,
Shanxi University, Taiyuan 030006, People's Republic of China.}
\affiliation{Collaborative Innovation Center of Extreme Optics, Shanxi University, Taiyuan 030006,
People's Republic of China.}
\author{Silvia Cardenas-Lopez}
\affiliation{Department of Physics, Columbia University, New York, NY 10027, USA.}
\author{Ana~Asenjo-Garcia}
\affiliation{Department of Physics, Columbia University, New York, NY 10027, USA.}
\author{Pablo Solano}
\affiliation{ Departamento de F\'isica, Facultad de Ciencias F\'isicas y Matem\'aticas, Universidad de Concepci\'on, Concepci\'on, Chile}
\affiliation{CIFAR Azrieli Global Scholars program, CIFAR, Toronto, Canada.}
\author{Luis A. Orozco} 
\affiliation{Joint Quantum Institute, Department of Physics and NIST, University of Maryland, College Park, MD 20742, USA.}
\author{Yanting Zhao}
\email{zhaoyt@sxu.edu.cn}
\affiliation{State Key Laboratory of Quantum Optics and Quantum Optics Devices, Institute of Laser Spectroscopy,
Shanxi University, Taiyuan 030006, People's Republic of China.}
\affiliation{Collaborative Innovation Center of Extreme Optics, Shanxi University, Taiyuan 030006,
People's Republic of China.}
\date{\today}
\begin{abstract} We study temporal oscillations, known as dynamical beats, developed by a propagating pulse due to its interaction with a near-resonant collective medium of $^{133}$Cs atoms randomly captured by a nanofiber-based optical lattice. A phenomenological theory provides an intuitive explanation and quantitative predictions, which are improved by an input-output theory considering multiple-scattering between the atoms. The results deepen our understanding of light propagation in waveguide QED, essential in time-frequency analysis and light engineering for probing, manipulating, and exploiting many-body quantum systems.
\end{abstract}
\maketitle
{\it Introduction.---} The study of electromagnetic pulse propagation through an absorbing medium has a long history in physics and engineering. Early studies by Sommerfeld \cite{Sommerfeld1914} and Brillouin \cite{Brillouin1914,Brillouin1960} showed the existence of what they called forerunners, later known as precursors~\cite{Oughstun1994,Macke09,Macke12,Macke13}, which are clearly observed when the rise-time of a pulse is shorter than the lifetime of the excited state of the resonant media \cite{Toyoda1997,Jeong2006,Jennewein2016,Jennewein2018}. Moreover, as the propagating pulse decays, it shows amplitude oscillations \cite{Lynch1960}, as observed by Lynch {\it et al.} in their early work on M{\"o}ssbauer spectroscopy and explained shortly after that  with a quantum theory by Harris \cite{Harris1961}. These oscillations are called dynamical beats (DB) \cite{burck1999}, most noticeably appearing in M{\"o}ssbauer experiments \cite{Hastings1991} but also present in the optical regime \cite{Crisp1970,CrispErr1970,Garrett1970}. DBs can be affected by the collective atomic response of the media, superradiantly speeding up the initial decay \cite{Adams2009,Svidzinsky2009,Rohlssbeger2010}, subradiantly slowing down the long time behavior, and altering the DB time evolution. In particular, the temporal structure of a guided pulse in an optical waveguide will significantly change upon interaction with an atomic ensemble, the centerpiece of waveguide quantum electrodynamics (QED) \cite{cardenas2023}.

\begin{figure}[t]
\begin{center}
\includegraphics[width=0.48\textwidth]{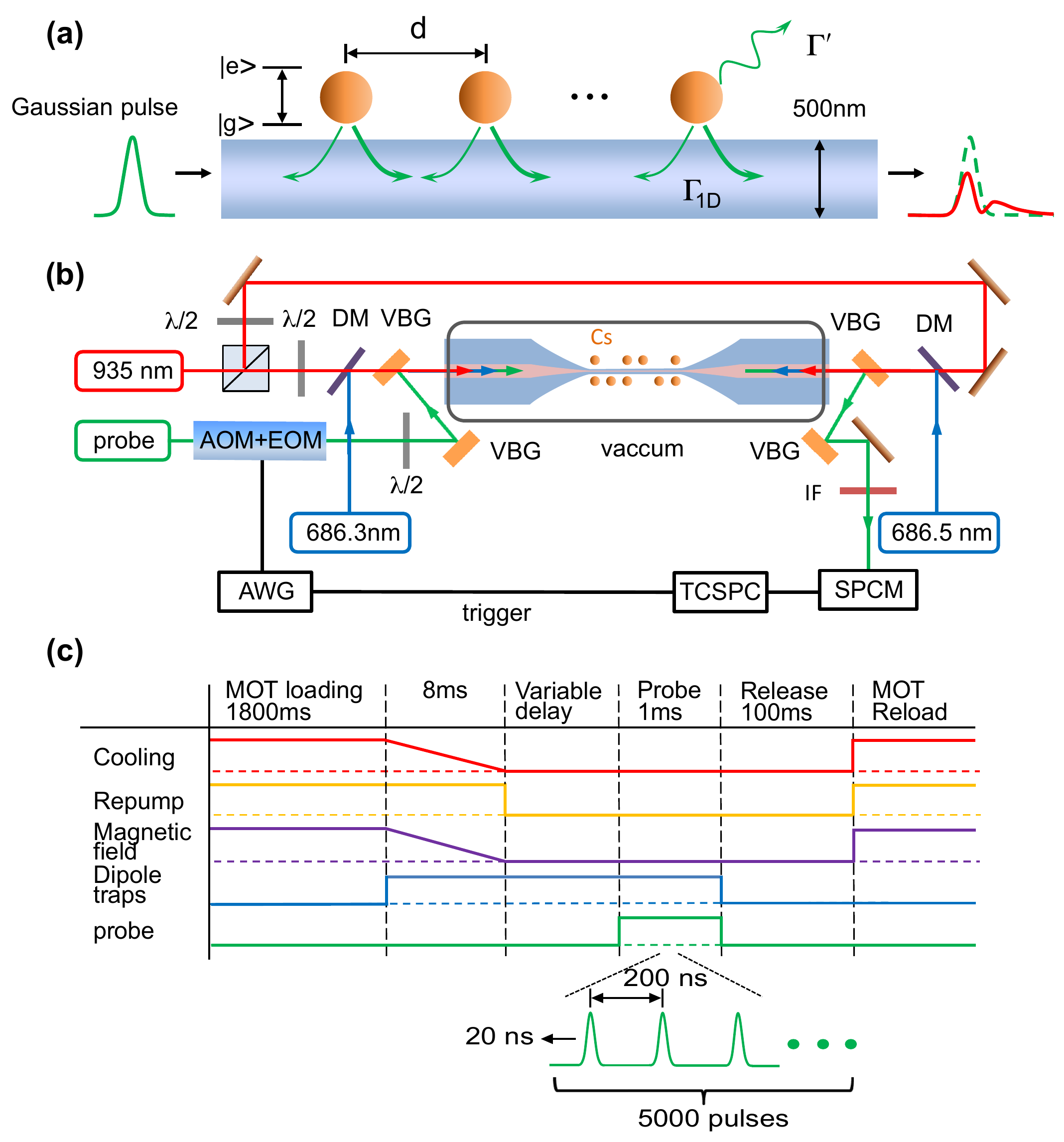}
\caption{ \textbf{(a)} The atomic array is trapped in the evanescent field of an optical lattice near the nanofiber surface. The lattice constant $d$ is half the wavelength of the trapping light. The temporal shape of a Gaussian input pulse (green) gets modified after propagating through an array of trapped atoms (red). {\textbf{(b)}} Schematic of the apparatus includes dichroic mirrors (DM), volume Bragg gratings (VBG), half wave plates $(\lambda/2)$ with the pulses coming from the left detecting them with single photon counter module (SPCM) on the right. The pulses are produced by  acousto- and electro-optical modulators (AOM, EOM) controlled by an electronic pulse from an arbitrary wave generator (AWG) that triggers the time correlator for single photon counters (TCSPC) for further data processing. {\textbf{(c)}} Time sequence for the experiment. The horizontal axis is not to scale.}
\vspace{-0.8cm}
\label{apparatus}
\end{center}
\end{figure}

Here, we experimentally and theoretically study what happens to a light pulse confined in a waveguide after crossing a medium consisting of an ordered array of atoms, as depicted by Fig.~\ref{apparatus}(a). In particular, we focus on the limit of resonant pulses with a temporal width shorter than the atomic lifetime. We can divide the resulting temporal behavior into two stages: at short times, while the pulse is still in the medium, there is a buildup of a macroscopic polarization (where the atoms are excited into a superposition state). After that, the temporal profile of the transmitted pulse is determined by the radiative decay of the macroscopic polarization (or collective atomic radiation)~\cite{cardenas2023,Pennetta2022a,Pennetta2022b,Kumlin2020}. A transmission dip, or zero, distinctively separates both regimes denoting the moment when the induced polarization amplitude matches the electric field amplitude of the input pulse but with the opposite phase. The versatility of waveguide QED is further enhanced with the quantitative understanding of its filtering properties in transmission. In designing the characteristics of the absorption and dispersion of the medium it is possible to shape a pulse in time and frequency, making the waveguide more than a transport element. 

{\it Experimental setup.---}
Hundreds of Cs atoms are optically confined around the surface of an optical nanofiber \cite{solano17d}. The nanofiber is 500 nm in diameter over a length of 5 mm fabricated from a standard optical fiber by flame brushing technique~\cite{hoffman14} (see Fig.~\ref{apparatus} {\bf(a,b)}). We use a two wavelengths optical dipole trap \cite{Vetsch2010}, operating close to the magic wavelengths to reduce the Stark shift of the trapped atoms \cite{Goban2012,Gobanerrata2012}. The calculated coupling of a trapped atom into the waveguide mode is $\Gamma_\text{1D}\approx 0.03 \Gamma'$, being $\Gamma'$ the emission rate into free space, for the operating transition $6S_{1/2}, F=4 \rightarrow 6P_{3/2}, F=5$. A magneto-optical trap captures Cs atoms from the residual gas in the vacuum chamber, which then fall into the dipole traps around the nanofiber. 
The lifetime of trapped atoms in the nanofiber optical lattice is about 8~ms, which is used to obtain a given optical depth, $OD$, by different delay times after loading into the nanofiber dipole trap. Although the $OD$ (measured by absorption spectroscopy) is reproducible, the exact location of the atoms within the periodic lattice is not. The influence of this disorder has been thoroughly studied in Ref.~\cite{cardenas2023} and it is not an issue in the present work. 

Figure~\ref{apparatus} {\bf(c)} shows the time sequence of the experiment using photon counting techniques. We produce the pulses with a combination of a fiber EOM and AOM that allows for shaping the pulses, changing their characteristic rise and fall times. We limit the length of the pulses to less than $1/\Gamma'$, the natural atomic lifetime. We refrain from using square pulses whose sharp edges can excite a superposition of the hyperfine excited state levels that complicate the interpretation of the measurements. An electronic reference pulse triggers the optical pulse and sets the zero time. A data acquisition card time-stamps the electronic pulses from the detectors for further data processing. The power of the excitation, small compared to the saturation intensity, corresponds to an average of $\approx0.4$ photons per pulse, guaranteeing less than two photons 94$\%$ of the time. For a given $OD$, we repeat the process of preparation, excitation, and measurement 5,000 times to obtain good statistics.

\begin{figure*}[t]
\centering
            \includegraphics[width=1.0\textwidth]{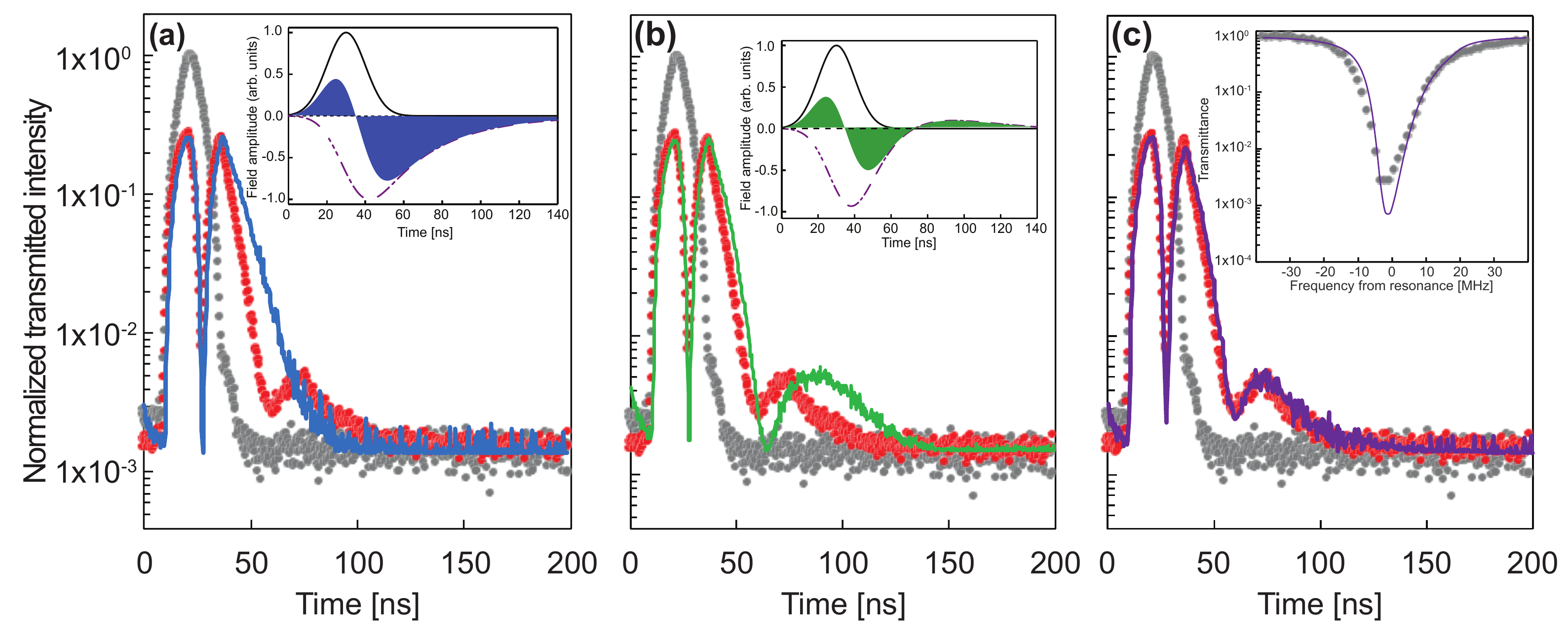}
            \caption{Time dependence of the transmitted intensity (red) for a 10~ns FWHM pulse propagating in a medium of OD=11.6, normalized by the peak intensity of the pulse without atoms (gray), in logarithmic scale. {\bf (a)}  The continuous blue line shows the results from the one-mode model of Eq.~(\ref{eq_single_mode}). The inset is the single mode theory field amplitude of the Gaussian input pulse (black), the atomic polarization (dashed purple), and the resulting sum, filled, producing one zero (blue). {\bf (b)} The continuous green line shows the results for the multimode model. The inset is the multimode theory field amplitude of the Gaussian input pulse (black), the atomic polarization (dashed purple), and the resulting sum, filled, producing two zeroes (green). {\bf (c)} The continuous purple line shows the results of using the broadened absorption with a sum of Lorentzians on Eqs.~\eqref{FT of Transmission} and~\eqref{eq:transmission} to obtain a fit (purple line) on the inset to the measurement gray dots. In all plots, $\Gamma'/2\pi$=5.2 MHz.}
            \vspace{-0.6cm}
            \label{Fig:2time_response}
\end{figure*}

Figure~\ref{Fig:2time_response} show the transmission for pulses with 10 ns full width at half maximum (FWHM). Gray points shows the transmitted intensity in the absence of atoms. The red dots show the transmitted pulse in the presence of atoms, where three peaks evidence the DBs. The three figures show the same data set but with different theoretical models, as we explain in the following sections.

{\it Phenomenological model.---}The primary physical mechanism underpinning the temporal modification of a pulse propagating through an atomic resonant medium is captured by considering a single electromagnetic mode and a macroscopic polarizable medium. The dynamics of the atomic polarization and the field are alike to those observed in cavity QED under pulse excitation \cite{Kaluzny1983,Mielke1997}, where the polarization of the atoms interferes with the electric field of the cavity. Similarly, in our case, the DBs are a consequence of the time-frequency response of the collective system coupled to the transmission waveguide. We consider a toy model motivated by the low intensity and bad-cavity limits of cavity QED, as the one developed by Carmichael {\it et al.} \cite{carmichael91}. The transmission through the system results from adding the electric field pulse $E_0(t)$ and the polarization field induced in the medium $p(t)$. The differential equation for the induced polarization is:
\begin{equation}
\frac{dp(t)}{dt}+\frac{\gamma}{2} p(t)= i\Omega(t),
\label{eq_single_mode}
\end{equation}
where $\Omega(t)$ is the field in units of Rabi frequency, and 
$\gamma=\Gamma'\left(1+ OD/2\right)$ is the atomic polarization decay constant. 

As shown in Fig.~\ref{Fig:2time_response}(a), this is a minimal model that captures the interplay between the pulse and the induced polarization: the induced polarization grows with a phase opposite to the external field, eventually causing a reduction in the transmission, which grows again when the external pulse turns off, and light is emitted solely from the depolarizing medium. The position of such local minimum observed in the transmission depends on the OD and the width and shape of the pulse. This behavior has been observed in cavity QED \cite{Mielke1997}. Local minima at later times occur when the input pulse is negligible, so they necessarily stem from interference effects that are not captured by this simple model. 

{\it Many-atoms theory.---} To develop a microscopic theory, one has to account for the many atoms coupled to the waveguide, and for the propagating nature of the guided mode. Generically, the transmitted field reads
\begin{equation}
E(t)=\frac{1}{2\pi}\int_{-\infty}^{\infty}\mathcal{T}(\omega)E_0(\omega)e^{-i\omega t}d\omega,
\label{FT of Transmission}
\end{equation}
where $E_0(\omega)$ is the Fourier transform of the input pulse, and $\mathcal{T}(\omega)$ is the transmission coefficient of the waveguide in the presence of atoms. 

We obtain the transmission coefficient via an input-output theory that enables us to write the (linear) response of the system in terms of $N$ collective atomic modes, where $N$ is the atom number~\cite{Asenjo2017, solano17a}. These modes, which can be either super- or sub-radiant, emerge from the atom-atom interactions mediated by the waveguide. They are found by diagonalizing the single-excitation sector of the non-Hermitian Hamiltonian 
\begin{equation}
\mathcal{H}_\text{1D}=-i \frac{\hbar\Gamma_\text{1D}}{2}\sum_{i,j=1}^N e^{ik d |i-j|} \hat{\sigma}^i_{eg}\hat{\sigma}^j_{ge}.\label{ham}
\end{equation}
Here $k$ is the guided-mode wavevector, $d$ is the inter-atomic separation, and $\hat{\sigma}^j_{ge}$ is the coherence operator for atom $j$. 

Plugging the steady-state solution of the Heisenberg-Langevin equations for the atomic coherences into an input-output equation for a monochromatic electromagnetic field, the transmission coefficient is found to be~\cite{Asenjo2017}
\begin{equation}
    T(\omega)=1-\frac{i\Gamma_\text{1D}}{2}\sum_{\xi=1}^N \frac{\eta_\xi}{\omega-\omega_0+i\Gamma'/2-\lambda_\xi}.
    \label{eq:response_func}
\end{equation}
Here, $\omega_0$ is the atomic resonance frequency and $\{\lambda_\xi\}$ are the (single-excitation) eigenvalues of $\mathcal{H}_\text{1D}$, which encode the frequency shifts and decay rates of the collective modes. The parameter $\eta_\xi=\sum_{n,m=1}^N v_{\xi,n}v_{\xi,m}e^{-ikd(n-m)}$ denotes the spatial overlap of the external field with the eigenvector $\textbf{v}_\xi$. The phenomenological expression for the transmission coefficient arising from Eq.~\eqref{eq_single_mode} is recovered when atoms are in the so-called ``mirror configuration'' (for which $kd$ is an integer multiple of $\pi$). In this configuration there is only a single superradiant mode and the Heisenberg-Langevin equations are identical to those of a collection of atoms in a bad cavity [see Supplementary Information (SI) ~\bibnote[SI]{See Supplementary Information for further details on the theory model, transmitted intensity in terms of collective modes and in the continuous limit, absorption sprectrum and data processing}]. 

The output field is the sum of the input field, $E_0(t)$, and the scattered field by the different collective modes. For an input pulse with peak-intensity $I_0$ and a Gaussian temporal envelope of variance $\sigma^2$, i.e., $f(t)=e^{-t^2/\sigma^2}/\sqrt{2\pi\sigma^2}$, the transmitted intensity reads~\bibnotemark[SI]
\begin{align}
    \frac{I_d(t)}{I_0}&=\bigg|f(t)e^{-i\Delta t}\nonumber\\&-\frac{\Gamma_{1D}}{4}\sum_{\xi=1}^N a_{\xi}e^{-(i\lambda_\xi+\Gamma'/2)t}\text{erfc}\left(\frac{b_\xi-t}{\sqrt{2}\sigma}\right)\bigg|^2,
    \label{eq:IntensityNmodes}
\end{align}
where erfc$(\cdot)$ is the complementary error function and $a_{\xi}$ and $b_{\xi}$ are expressions given in the SI~\bibnotemark[SI]. 

The first zero in Fig.~\ref{Fig:2time_response} appears when the two terms in the above equation cancel each other. In the limit of large optical depth (i.e., for $OD\equiv 2N\Gamma_\text{1D}/\Gamma'\gg 1$), the first zero appears at short times compared to the natural lifetime $1/\Gamma'$. In the crude limit of considering only the most superradiant mode (with decay rate $\sim N\Gamma_\text{1D}$), a Taylor expansion of the above expression allows us to find the time for the first zero as $\Gamma'\tau_\text{zero}\simeq 4/OD$. The subsequent zeros, i.e., the dynamical beats, arise from interference between different collective atomic modes. However, the phenomenology cannot be simply attributed to a few dominant modes, it is truly a multi-mode feature. 

In the large atom limit, the transmission coefficient in Eq.~\eqref{eq:response_func} converges to that of a continuous dielectric medium~\cite{cardenas2023}, i.e., 
\begin{equation}
    \mathcal{T}(\omega)=\text{exp}\left(-\frac{i N \Gamma_{\rm{1D}}}{2}\frac{1}{\omega-\omega_0+i\Gamma'/2}\right).
    \label{eq:transmission}
\end{equation}
The above transmission coefficient allows for a semi-analytical expression of the transmitted pulse intensity. The details are fully developed in Ref.~\cite{cardenas2023}, where we investigate the transport of broadband square photon pulses in waveguide QED, and the calculation is inspired by Ref.~\cite{Harris1961}. Here we adapt these results to Gaussian input pulses.  After performing the integral in frequency, the output intensity can be written as series of Bessel functions, i.e.,
\begin{equation}
\frac{I_d(t)}{I_0}=e^{-\Gamma' t}\left|\sum_{m=-\infty}^\infty A_m\left(\sqrt{\frac{t}{OD\Gamma'}}\right)^m J_m\left(\sqrt{ OD \Gamma' t}\right)\right|^2,
\label{eq:FinalIntensity}
\end{equation}
where the expressions for the coefficients $A_m$ -- which depend on the pulse parameters -- are given in the SI~\bibnotemark[SI]. 

As can be seen from this expression, the optical depth determines the timescale of the dynamics. Moreover, while our Eq.~\eqref{ham} refers to an ordered array, the convergence of the transmission coefficient to that of a continuous medium for large atom number indicates that position disorder is irrelevant for our results. This last point is in contrast with what occurs for reflection, which is only observable in periodic arrays~\cite{Deutsch1995,Corzo2016,Sorensen2016}.

{\it Experimental analysis.---} 
The transmitted pulse, shown in Fig.~\ref{Fig:2time_response}, has three peaks and two valleys. The first valley is reproduced in {\bf (a)} by the continuous blue line of the single mode macroscopic theory as in Eq.~(\ref{eq_single_mode}), which only considers a single electromagnetic mode and a polarizable medium. It shows the interference between the two with a significant decrease in the transmission at around 40 ns (see inset for the field and atomic polarization amplitude). 

The second valley is reproduced by the many-atom theory, as shown in {\bf(b)}. This theory also captures the relatively fast decay of the data on the way to the second valley. To produce the theory curve, we solve the Heisenberg-Langevin equations for the atomic coherences for the input drive (which we extract from the experiment). We then compute the transmitted field via input-output equations. More details are provided in the SI~\bibnotemark[SI].

The amplitude of the third peak is correctly predicted by the many-atom theory but it emerges later in the calculation than in the measurement. An exploration of the numerical simulations shows that this discrepancy is due to line broadening. To account for this disparity, we model the complex transmission coefficient of the asymmetrically and inhomogeneously broadened atomic medium as the product of frequency-displaced transmission coefficients of the form of Eq.~(\ref{eq:transmission}), each with a different number of atoms~\bibnotemark[SI]. The number of atoms corresponding to each shifted transmission is sampled from a log-normal distribution \cite{Patterson2018}, characteristic of random processes bounded on one side, such as position-dependent, positive-only light shifts. The model reproduces to a good approximation the measured transmission spectrum shown in the inset of {\bf(c)}. Considering the modeled transmission coefficient, we produce the purple line from the square of Eq.~(\ref{FT of Transmission}), in excellent agreement with the pulse measurements. 

\begin{figure}[t]
\begin{center}
\includegraphics[width=0.45\textwidth]{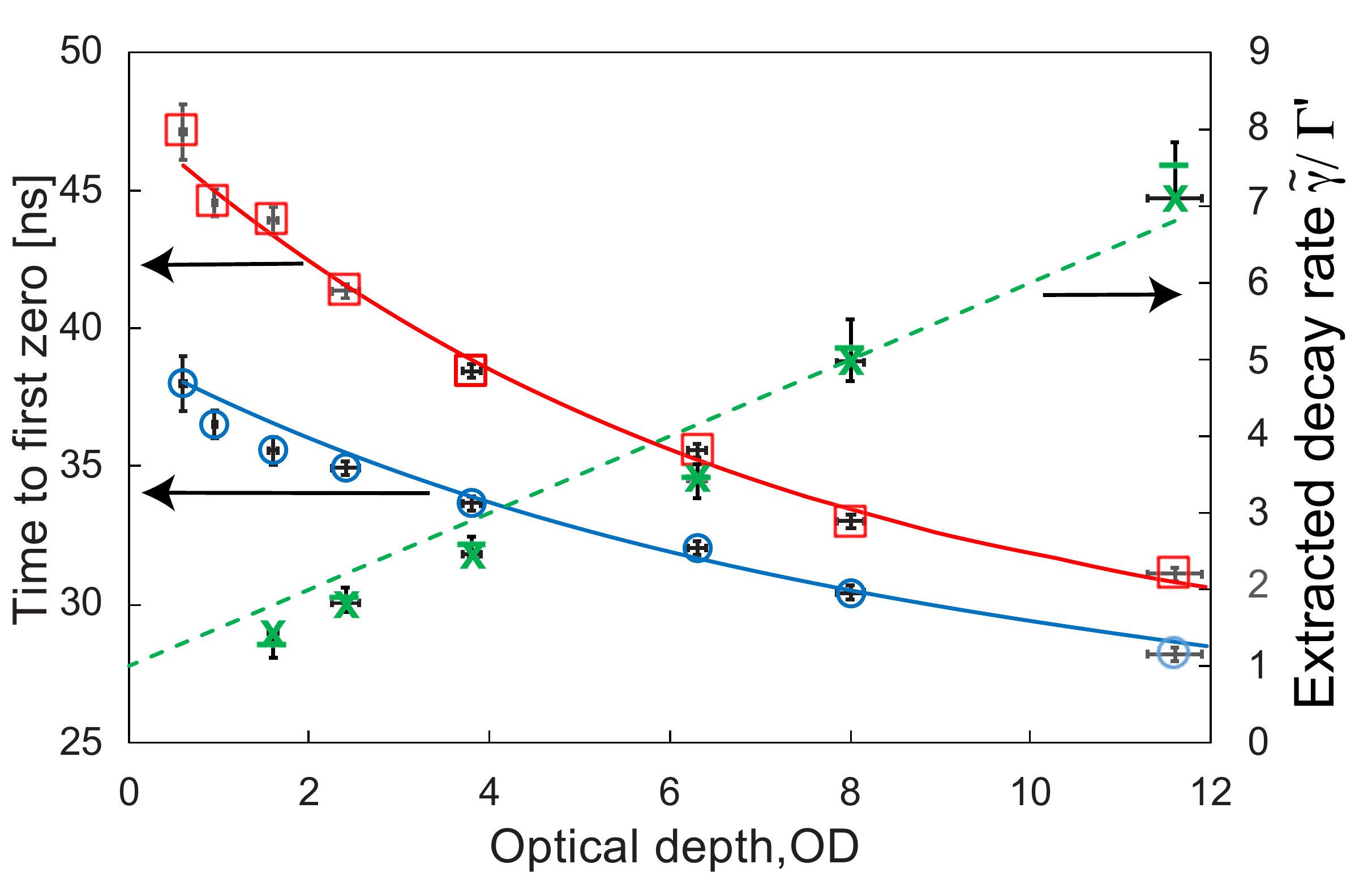}
\caption{Left: Evolution of the first zero with on-resonance excitation (valley) of the transmission as a function of the OD for two different pulse widths. Blue circles correspond to a FWHM of 10~ns and the red squares for a FWHM of 13~ns. The corresponding blue and red continuous lines indicate the prediction of the full theory. Right: Extracted decay rate (green) from the fall of the second peak when the input is a 10~ns pulse (bars), 13~ns (exes).  The dashed green line is 1+OD/2, the expected value from the theory. The error bars are the same for the two sets of data, but only the horizontal or the vertical are plotted for clarity.}
\vspace{-1cm}
\label{Fig:3first-zero-slope}
\end{center}
\end{figure}

Figure~\ref{Fig:3first-zero-slope} shows the location of the first zero and the slope of the second fall from a series of pulse transmission measurements, as a function of the optical depth. Two sets of data, corresponding to pulses of 10 ns (blue circle) and 13 ns FWHM (red square) show the delay of the first zero as a function of $OD$. The theory curves (red, blue), as obtained from the many-atom theory, agree with the experimental data. The curves are produced with  a single adjustable parameter, the offset from the peak of the input pulse (1.1 ns optimizes the fit). As expected, the time of the first zero decreases with increasing optical depth. The dependence on the pulse width becomes less relevant as the optical depth increases, in agreement with our theoretical model (which predicts that the first zero occurs  at a time  $\Gamma'\tau_\text{zero}\simeq 4/OD$ after the input pulse center in the large OD limit). However, while the theory captures well the scaling with the optical depth, the approximation that yields a simple expression is only in qualitative (but not quantitative) agreement, as the $OD$ in the experiment is not large enough. 

The right axis of Fig. \ref{Fig:3first-zero-slope} shows an effective decay rate obtained from a fit of the second fall on the transmitted pulses of 10 ns (bars) and 13 ns (exes). Both data sets have statistical error bars, and only one of them is plotted for clarity. The extracted numbers  fall in the same range. The rate value seems to only be dependent on the $OD$. Once the absorption is large enough, the decay
rate grows almost linearly with the increased $OD$. An enhanced decay rate arises because the superradiant modes in Eq.~\eqref{eq:IntensityNmodes} dominate the signal at early times. By assuming that the decay rate is determined by these modes we obtain $\tilde{\gamma}\sim \Gamma'(1+OD/2)$ ~\bibnotemark[SI], which is consistent with the experimental results. The decay rate changes at later times, but experimental noise prevents us from accessing this regime.

{\it Discussion.---}
We measured the emergence of oscillations in the temporal behavior of a pulse propagating through a resonant atomic medium. After contrasting different models to the measurements, we find that the position of the first valley and the first two peaks are mainly insensitive to the absorption details near resonance. This is because the transmittance is practically zero near resonance. On the other hand, the tails of the spectral distribution contribute much more to the second valley and third peak. Although good from a statistical point of view, the fits to the measured absorption spectrum present deviations in the tails. Combinations of the many magnetic sublevels, the position-dependent light-shifts induced by the trap, the remaining heating of the trapped atoms, and the nanofiber torsional modes \cite{Su2022} can contribute to this issue. 

Dynamical beats are sensitive to both the amplitude and the phase of the (complex) transmission coefficient. However, extracting this information from the current measurement is not possible. To do so, one would need frequency-dependent homodyne measurements. Nonetheless, knowing the main shift sources allows us to develop an approximate model for the complex transmission coefficient of the system that yields a good agreement with the experiment.

In the waveguide QED regime amenable to nanofiber experiments, a quantum description for the atomic response is not needed to understand DBs. Although our approach comes from a microscopic description~\cite{cardenas2023}, in the many-atom limit, the results agree perfectly with those predicted for a continuous medium.  This result shows that a classical description based on the linear transmission coefficient (for a continuous medium) and input electric field is sufficient in many waveguide QED transmittance measurements~\cite{Pennetta2022a,Pennetta2022b,Kumlin2020}, in particular when both the saturation and the ratio $\Gamma_\text{1D}/\Gamma'$ are low.

One could extend the study to larger intensities, with many excitations, to evaluate how the quantum nonlinear responses of the system emerge in a transient regime. One could also study DBs for more complex atomic structures, such as three-level systems, where electromagnetically induced transparency can further separate an optical precursor from the main pulse~\cite{Wei2009, cardenas2023}. In such a case, the transmission coefficient found for a continuous medium would need to be modified, and further experiments would be required as the parameter space broadens.

{\it Conclusions.---}
We experimentally and theoretically study the problem of pulse propagation through a resonant media, showing the emergence of dynamical beats in waveguide QED. We focus on the limit of pulses with temporal widths below the atomic lifetime $1/\Gamma'$ and below the atomic saturation intensity. An optical nanofiber guides a single spatial mode of light propagating through a periodic ensemble of optically trapped atoms, providing a valuable platform to study the phenomenon. In order to understand the most relevant features in the emergence of dynamical beats, we propose two theoretical models: a simplified one, which explains the main structures of the transmitted light pulse, and an effective one based on macroscopic electrodynamics, which can be microscopically derived from a multiple-scattering input-output theory.  The latter correctly describes the transmitted pulse oscillations at longer times. Moreover, we find that the off-resonance structure of the transmission coefficient of the sample becomes relevant to the long-time beating behavior of the output pulse. Finally, we discuss that an accurate description of the transmitted pulse and the dynamical beats in the low-intensity limit does not require knowledge of the atomic states of the system, and classical field calculations are sufficient.
Our results provide insights into the key factors determining the emergence of dynamical beats, a relevant effect for light-based communication and information processing protocols that relies on sending pulses to interconnect resonant samples or quantum emitters. The control of the temporal frequency characteristics of the pulse given by the atomic properties such as the $OD$ opens the field to the construction of tailored transmission filters. 

{\it Acknowledgements.---}
P.S. is a CIFAR Azrieli Global Scholar in the Quantum Information Science Program. This work was supported by the National Key Research and Development Program of China (No. 2022YFA1404201), National Natural Science Foundation of China (Nos. 12034012, 12274272, 61827824, 62105191, 12074231), Fundamental Research Program of Shanxi Province(20210302124537), ``1331 KSC'', PCSIRT (No. IRT\_17R70), 111 Project (No. D18001), CONICYT-PAI grant 77190033, and FONDECYT grant N$^{\circ}$ 11200192 from Chile.  We also gratefully acknowledge support from the Air Force Office of Scientific Research through their Young Investigator Prize (grant No.~21RT0751), the National Science Foundation through their CAREER Award (No. 2047380), the A. P. Sloan foundation, and the David and Lucile Packard foundation. S. C.-L. acknowledges additional support from the Chien-Shiung Wu Family Foundation.
\vspace{-0.70cm}

\bibliography{Aonf,Torsion,AAMOPreview}

\begin{thebibliography}{42}%
\makeatletter
\providecommand \@ifxundefined [1]{%
 \@ifx{#1\undefined}
}%
\providecommand \@ifnum [1]{%
 \ifnum #1\expandafter \@firstoftwo
 \else \expandafter \@secondoftwo
 \fi
}%
\providecommand \@ifx [1]{%
 \ifx #1\expandafter \@firstoftwo
 \else \expandafter \@secondoftwo
 \fi
}%
\providecommand \natexlab [1]{#1}%
\providecommand \enquote  [1]{``#1''}%
\providecommand \bibnamefont  [1]{#1}%
\providecommand \bibfnamefont [1]{#1}%
\providecommand \citenamefont [1]{#1}%
\providecommand \href@noop [0]{\@secondoftwo}%
\providecommand \href [0]{\begingroup \@sanitize@url \@href}%
\providecommand \@href[1]{\@@startlink{#1}\@@href}%
\providecommand \@@href[1]{\endgroup#1\@@endlink}%
\providecommand \@sanitize@url [0]{\catcode `\\12\catcode `\$12\catcode
  `\&12\catcode `\#12\catcode `\^12\catcode `\_12\catcode `\%12\relax}%
\providecommand \@@startlink[1]{}%
\providecommand \@@endlink[0]{}%
\providecommand \url  [0]{\begingroup\@sanitize@url \@url }%
\providecommand \@url [1]{\endgroup\@href {#1}{\urlprefix }}%
\providecommand \urlprefix  [0]{URL }%
\providecommand \Eprint [0]{\href }%
\providecommand \doibase [0]{http://dx.doi.org/}%
\providecommand \selectlanguage [0]{\@gobble}%
\providecommand \bibinfo  [0]{\@secondoftwo}%
\providecommand \bibfield  [0]{\@secondoftwo}%
\providecommand \translation [1]{[#1]}%
\providecommand \BibitemOpen [0]{}%
\providecommand \bibitemStop [0]{}%
\providecommand \bibitemNoStop [0]{.\EOS\space}%
\providecommand \EOS [0]{\spacefactor3000\relax}%
\providecommand \BibitemShut  [1]{\csname bibitem#1\endcsname}%
\let\auto@bib@innerbib\@empty
\bibitem [{\citenamefont {Sommerfeld}(1914)}]{Sommerfeld1914}%
  \BibitemOpen
  \bibfield  {author} {\bibinfo {author} {\bibfnamefont {A.}~\bibnamefont
  {Sommerfeld}},\ }\href@noop {} {\bibfield  {journal} {\bibinfo  {journal}
  {Ann. Phys. (Leipzig)}\ }\textbf {\bibinfo {volume} {44}},\ \bibinfo {pages}
  {177} (\bibinfo {year} {1914})}\BibitemShut {NoStop}%
\bibitem [{\citenamefont {Brillouin}(1914)}]{Brillouin1914}%
  \BibitemOpen
  \bibfield  {author} {\bibinfo {author} {\bibfnamefont {L.}~\bibnamefont
  {Brillouin}},\ }\href@noop {} {\bibfield  {journal} {\bibinfo  {journal}
  {Ann. Phys. (Leipzig)}\ }\textbf {\bibinfo {volume} {44}},\ \bibinfo {pages}
  {203} (\bibinfo {year} {1914})}\BibitemShut {NoStop}%
\bibitem [{\citenamefont {Brillouin}(1960)}]{Brillouin1960}%
  \BibitemOpen
  \bibfield  {author} {\bibinfo {author} {\bibfnamefont {L.}~\bibnamefont
  {Brillouin}},\ }\href@noop {} {\emph {\bibinfo {title} {Wave Propagattion and
  Group Velocity}}}\ (\bibinfo  {publisher} {Academic},\ \bibinfo {year}
  {1960})\BibitemShut {NoStop}%
\bibitem [{\citenamefont {Oughstun}\ and\ \citenamefont
  {Sherman}(1994)}]{Oughstun1994}%
  \BibitemOpen
  \bibfield  {author} {\bibinfo {author} {\bibfnamefont {K.~E.}\ \bibnamefont
  {Oughstun}}\ and\ \bibinfo {author} {\bibfnamefont {G.~C.}\ \bibnamefont
  {Sherman}},\ }\href@noop {} {\emph {\bibinfo {title} {Electromagnetic Pulse
  Propagation in Causal Dielectrics}}}\ (\bibinfo  {publisher} {Springer
  Verlag},\ \bibinfo {year} {1994})\BibitemShut {NoStop}%
\bibitem [{\citenamefont {Macke}\ and\ \citenamefont
  {S\'egard}(2009)}]{Macke09}%
  \BibitemOpen
  \bibfield  {author} {\bibinfo {author} {\bibfnamefont {B.}~\bibnamefont
  {Macke}}\ and\ \bibinfo {author} {\bibfnamefont {B.}~\bibnamefont
  {S\'egard}},\ }\href {\doibase 10.1103/PhysRevA.80.011803} {\bibfield
  {journal} {\bibinfo  {journal} {Phys. Rev. A}\ }\textbf {\bibinfo {volume}
  {80}},\ \bibinfo {pages} {011803} (\bibinfo {year} {2009})}\BibitemShut
  {NoStop}%
\bibitem [{\citenamefont {Macke}\ and\ \citenamefont
  {S\'egard}(2012)}]{Macke12}%
  \BibitemOpen
  \bibfield  {author} {\bibinfo {author} {\bibfnamefont {B.}~\bibnamefont
  {Macke}}\ and\ \bibinfo {author} {\bibfnamefont {B.}~\bibnamefont
  {S\'egard}},\ }\href {\doibase 10.1103/PhysRevA.86.013837} {\bibfield
  {journal} {\bibinfo  {journal} {Phys. Rev. A}\ }\textbf {\bibinfo {volume}
  {86}},\ \bibinfo {pages} {013837} (\bibinfo {year} {2012})}\BibitemShut
  {NoStop}%
\bibitem [{\citenamefont {Macke}\ and\ \citenamefont
  {S\'egard}(2013)}]{Macke13}%
  \BibitemOpen
  \bibfield  {author} {\bibinfo {author} {\bibfnamefont {B.}~\bibnamefont
  {Macke}}\ and\ \bibinfo {author} {\bibfnamefont {B.}~\bibnamefont
  {S\'egard}},\ }\href {\doibase 10.1103/PhysRevA.87.043830} {\bibfield
  {journal} {\bibinfo  {journal} {Phys. Rev. A}\ }\textbf {\bibinfo {volume}
  {87}},\ \bibinfo {pages} {043830} (\bibinfo {year} {2013})}\BibitemShut
  {NoStop}%
\bibitem [{\citenamefont {Toyoda}\ \emph {et~al.}(1997)\citenamefont {Toyoda},
  \citenamefont {Takahashi}, \citenamefont {Ishikawa},\ and\ \citenamefont
  {Yabuzaki}}]{Toyoda1997}%
  \BibitemOpen
  \bibfield  {author} {\bibinfo {author} {\bibfnamefont {K.}~\bibnamefont
  {Toyoda}}, \bibinfo {author} {\bibfnamefont {Y.}~\bibnamefont {Takahashi}},
  \bibinfo {author} {\bibfnamefont {K.}~\bibnamefont {Ishikawa}}, \ and\
  \bibinfo {author} {\bibfnamefont {T.}~\bibnamefont {Yabuzaki}},\ }\href
  {\doibase 10.1103/PhysRevA.56.1564} {\bibfield  {journal} {\bibinfo
  {journal} {Phys. Rev. A}\ }\textbf {\bibinfo {volume} {56}},\ \bibinfo
  {pages} {1564} (\bibinfo {year} {1997})}\BibitemShut {NoStop}%
\bibitem [{\citenamefont {Jeong}\ \emph {et~al.}(2006)\citenamefont {Jeong},
  \citenamefont {Dawes},\ and\ \citenamefont {Gauthier}}]{Jeong2006}%
  \BibitemOpen
  \bibfield  {author} {\bibinfo {author} {\bibfnamefont {H.}~\bibnamefont
  {Jeong}}, \bibinfo {author} {\bibfnamefont {A.~M.~C.}\ \bibnamefont {Dawes}},
  \ and\ \bibinfo {author} {\bibfnamefont {D.~J.}\ \bibnamefont {Gauthier}},\
  }\href {\doibase 10.1103/PhysRevLett.96.143901} {\bibfield  {journal}
  {\bibinfo  {journal} {Phys. Rev. Lett.}\ }\textbf {\bibinfo {volume} {96}},\
  \bibinfo {pages} {143901} (\bibinfo {year} {2006})}\BibitemShut {NoStop}%
\bibitem [{\citenamefont {Jennewein}\ \emph {et~al.}(2016)\citenamefont
  {Jennewein}, \citenamefont {Besbes}, \citenamefont {Schilder}, \citenamefont
  {Jenkins}, \citenamefont {Sauvan}, \citenamefont {Ruostekoski}, \citenamefont
  {Greffet}, \citenamefont {Sortais},\ and\ \citenamefont
  {Browaeys}}]{Jennewein2016}%
  \BibitemOpen
  \bibfield  {author} {\bibinfo {author} {\bibfnamefont {S.}~\bibnamefont
  {Jennewein}}, \bibinfo {author} {\bibfnamefont {M.}~\bibnamefont {Besbes}},
  \bibinfo {author} {\bibfnamefont {N.~J.}\ \bibnamefont {Schilder}}, \bibinfo
  {author} {\bibfnamefont {S.~D.}\ \bibnamefont {Jenkins}}, \bibinfo {author}
  {\bibfnamefont {C.}~\bibnamefont {Sauvan}}, \bibinfo {author} {\bibfnamefont
  {J.}~\bibnamefont {Ruostekoski}}, \bibinfo {author} {\bibfnamefont {J.-J.}\
  \bibnamefont {Greffet}}, \bibinfo {author} {\bibfnamefont {Y.~R.~P.}\
  \bibnamefont {Sortais}}, \ and\ \bibinfo {author} {\bibfnamefont
  {A.}~\bibnamefont {Browaeys}},\ }\href {\doibase
  10.1103/PhysRevLett.116.233601} {\bibfield  {journal} {\bibinfo  {journal}
  {Phys. Rev. Lett.}\ }\textbf {\bibinfo {volume} {116}},\ \bibinfo {pages}
  {233601} (\bibinfo {year} {2016})}\BibitemShut {NoStop}%
\bibitem [{\citenamefont {Jennewein}\ \emph {et~al.}(2018)\citenamefont
  {Jennewein}, \citenamefont {Brossard}, \citenamefont {Sortais}, \citenamefont
  {Browaeys}, \citenamefont {Cheinet}, \citenamefont {Robert},\ and\
  \citenamefont {Pillet}}]{Jennewein2018}%
  \BibitemOpen
  \bibfield  {author} {\bibinfo {author} {\bibfnamefont {S.}~\bibnamefont
  {Jennewein}}, \bibinfo {author} {\bibfnamefont {L.}~\bibnamefont {Brossard}},
  \bibinfo {author} {\bibfnamefont {Y.~R.~P.}\ \bibnamefont {Sortais}},
  \bibinfo {author} {\bibfnamefont {A.}~\bibnamefont {Browaeys}}, \bibinfo
  {author} {\bibfnamefont {P.}~\bibnamefont {Cheinet}}, \bibinfo {author}
  {\bibfnamefont {J.}~\bibnamefont {Robert}}, \ and\ \bibinfo {author}
  {\bibfnamefont {P.}~\bibnamefont {Pillet}},\ }\href {\doibase
  10.1103/PhysRevA.97.053816} {\bibfield  {journal} {\bibinfo  {journal} {Phys.
  Rev. A}\ }\textbf {\bibinfo {volume} {97}},\ \bibinfo {pages} {053816}
  (\bibinfo {year} {2018})}\BibitemShut {NoStop}%
\bibitem [{\citenamefont {Lynch}\ \emph {et~al.}(1960)\citenamefont {Lynch},
  \citenamefont {Holland},\ and\ \citenamefont {Hamermesh}}]{Lynch1960}%
  \BibitemOpen
  \bibfield  {author} {\bibinfo {author} {\bibfnamefont {F.~J.}\ \bibnamefont
  {Lynch}}, \bibinfo {author} {\bibfnamefont {R.~E.}\ \bibnamefont {Holland}},
  \ and\ \bibinfo {author} {\bibfnamefont {M.}~\bibnamefont {Hamermesh}},\
  }\href {\doibase 10.1103/PhysRev.120.513} {\bibfield  {journal} {\bibinfo
  {journal} {Phys. Rev.}\ }\textbf {\bibinfo {volume} {120}},\ \bibinfo {pages}
  {513} (\bibinfo {year} {1960})}\BibitemShut {NoStop}%
\bibitem [{\citenamefont {Harris}(1961)}]{Harris1961}%
  \BibitemOpen
  \bibfield  {author} {\bibinfo {author} {\bibfnamefont {S.~M.}\ \bibnamefont
  {Harris}},\ }\href {\doibase 10.1103/PhysRev.124.1178} {\bibfield  {journal}
  {\bibinfo  {journal} {Phys. Rev.}\ }\textbf {\bibinfo {volume} {124}},\
  \bibinfo {pages} {1178} (\bibinfo {year} {1961})}\BibitemShut {NoStop}%
\bibitem [{\citenamefont {van B{\"u}rck}(1999)}]{burck1999}%
  \BibitemOpen
  \bibfield  {author} {\bibinfo {author} {\bibfnamefont {U.}~\bibnamefont {van
  B{\"u}rck}},\ }\href {\doibase 10.1023/A:1017080008712} {\bibfield  {journal}
  {\bibinfo  {journal} {Hyperfine Interactions}\ }\textbf {\bibinfo {volume}
  {123}},\ \bibinfo {pages} {483} (\bibinfo {year} {1999})}\BibitemShut
  {NoStop}%
\bibitem [{\citenamefont {Hastings}\ \emph {et~al.}(1991)\citenamefont
  {Hastings}, \citenamefont {Siddons}, \citenamefont {van B\"urck},
  \citenamefont {Hollatz},\ and\ \citenamefont {Bergmann}}]{Hastings1991}%
  \BibitemOpen
  \bibfield  {author} {\bibinfo {author} {\bibfnamefont {J.~B.}\ \bibnamefont
  {Hastings}}, \bibinfo {author} {\bibfnamefont {D.~P.}\ \bibnamefont
  {Siddons}}, \bibinfo {author} {\bibfnamefont {U.}~\bibnamefont {van
  B\"urck}}, \bibinfo {author} {\bibfnamefont {R.}~\bibnamefont {Hollatz}}, \
  and\ \bibinfo {author} {\bibfnamefont {U.}~\bibnamefont {Bergmann}},\ }\href
  {\doibase 10.1103/PhysRevLett.66.770} {\bibfield  {journal} {\bibinfo
  {journal} {Phys. Rev. Lett.}\ }\textbf {\bibinfo {volume} {66}},\ \bibinfo
  {pages} {770} (\bibinfo {year} {1991})}\BibitemShut {NoStop}%
\bibitem [{\citenamefont {Crisp}(1970{\natexlab{a}})}]{Crisp1970}%
  \BibitemOpen
  \bibfield  {author} {\bibinfo {author} {\bibfnamefont {M.~D.}\ \bibnamefont
  {Crisp}},\ }\href {\doibase 10.1103/PhysRevA.1.1604} {\bibfield  {journal}
  {\bibinfo  {journal} {Phys. Rev. A}\ }\textbf {\bibinfo {volume} {1}},\
  \bibinfo {pages} {1604} (\bibinfo {year} {1970}{\natexlab{a}})}\BibitemShut
  {NoStop}%
\bibitem [{\citenamefont {Crisp}(1970{\natexlab{b}})}]{CrispErr1970}%
  \BibitemOpen
  \bibfield  {author} {\bibinfo {author} {\bibfnamefont {M.~D.}\ \bibnamefont
  {Crisp}},\ }\href {\doibase 10.1103/PhysRevA.2.2172.2} {\bibfield  {journal}
  {\bibinfo  {journal} {Phys. Rev. A}\ }\textbf {\bibinfo {volume} {2}},\
  \bibinfo {pages} {2172} (\bibinfo {year} {1970}{\natexlab{b}})}\BibitemShut
  {NoStop}%
\bibitem [{\citenamefont {Garrett}\ and\ \citenamefont
  {McCumber}(1970)}]{Garrett1970}%
  \BibitemOpen
  \bibfield  {author} {\bibinfo {author} {\bibfnamefont {C.~G.~B.}\
  \bibnamefont {Garrett}}\ and\ \bibinfo {author} {\bibfnamefont {D.~E.}\
  \bibnamefont {McCumber}},\ }\href {\doibase 10.1103/PhysRevA.1.305}
  {\bibfield  {journal} {\bibinfo  {journal} {Phys. Rev. A}\ }\textbf {\bibinfo
  {volume} {1}},\ \bibinfo {pages} {305} (\bibinfo {year} {1970})}\BibitemShut
  {NoStop}%
\bibitem [{\citenamefont {Adams}(2009)}]{Adams2009}%
  \BibitemOpen
  \bibfield  {author} {\bibinfo {author} {\bibfnamefont {B.}~\bibnamefont
  {Adams}},\ }\href {\doibase 10.1080/09500340903199921} {\bibfield  {journal}
  {\bibinfo  {journal} {Journal of Modern Optics}\ }\textbf {\bibinfo {volume}
  {56}},\ \bibinfo {pages} {1974} (\bibinfo {year} {2009})}\BibitemShut
  {NoStop}%
\bibitem [{\citenamefont {Svidzinsky}\ and\ \citenamefont
  {Scully}(2009)}]{Svidzinsky2009}%
  \BibitemOpen
  \bibfield  {author} {\bibinfo {author} {\bibfnamefont {A.~A.}\ \bibnamefont
  {Svidzinsky}}\ and\ \bibinfo {author} {\bibfnamefont {M.~O.}\ \bibnamefont
  {Scully}},\ }\href {\doibase https://doi.org/10.1016/j.optcom.2009.04.011}
  {\bibfield  {journal} {\bibinfo  {journal} {Optics Communications}\ }\textbf
  {\bibinfo {volume} {282}},\ \bibinfo {pages} {2894} (\bibinfo {year}
  {2009})}\BibitemShut {NoStop}%
\bibitem [{\citenamefont {{Ralf R{\"o}hlsberger and Kai Schlage and Balaram
  Sahoo and Sebastien Couet and Rudolf R{\"u}ffer }}(2010)}]{Rohlssbeger2010}%
  \BibitemOpen
  \bibfield  {author} {\bibinfo {author} {\bibnamefont {{Ralf R{\"o}hlsberger
  and Kai Schlage and Balaram Sahoo and Sebastien Couet and Rudolf R{\"u}ffer
  }}},\ }\href {https://www.science.org/doi/abs/10.1126/science.1187770}
  {\bibfield  {journal} {\bibinfo  {journal} {Science}\ }\textbf {\bibinfo
  {volume} {328}},\ \bibinfo {pages} {1248} (\bibinfo {year}
  {2010})}\BibitemShut {NoStop}%
\bibitem [{\citenamefont {Cardenas-Lopez}\ \emph {et~al.}(2023)\citenamefont
  {Cardenas-Lopez}, \citenamefont {Solano}, \citenamefont {Orozco},\ and\
  \citenamefont {Asenjo-Garcia}}]{cardenas2023}%
  \BibitemOpen
  \bibfield  {author} {\bibinfo {author} {\bibfnamefont {S.}~\bibnamefont
  {Cardenas-Lopez}}, \bibinfo {author} {\bibfnamefont {P.}~\bibnamefont
  {Solano}}, \bibinfo {author} {\bibfnamefont {L.~A.}\ \bibnamefont {Orozco}},
  \ and\ \bibinfo {author} {\bibfnamefont {A.}~\bibnamefont {Asenjo-Garcia}},\
  }\href {\doibase 10.1103/PhysRevResearch.5.013133} {\bibfield  {journal}
  {\bibinfo  {journal} {Phys. Rev. Res.}\ }\textbf {\bibinfo {volume} {5}},\
  \bibinfo {pages} {013133} (\bibinfo {year} {2023})}\BibitemShut {NoStop}%
\bibitem [{\citenamefont {Pennetta}\ \emph
  {et~al.}(2022{\natexlab{a}})\citenamefont {Pennetta}, \citenamefont {Blaha},
  \citenamefont {Johnson}, \citenamefont {Lechner}, \citenamefont
  {Schneeweiss}, \citenamefont {Volz},\ and\ \citenamefont
  {Rauschenbeutel}}]{Pennetta2022a}%
  \BibitemOpen
  \bibfield  {author} {\bibinfo {author} {\bibfnamefont {R.}~\bibnamefont
  {Pennetta}}, \bibinfo {author} {\bibfnamefont {M.}~\bibnamefont {Blaha}},
  \bibinfo {author} {\bibfnamefont {A.}~\bibnamefont {Johnson}}, \bibinfo
  {author} {\bibfnamefont {D.}~\bibnamefont {Lechner}}, \bibinfo {author}
  {\bibfnamefont {P.}~\bibnamefont {Schneeweiss}}, \bibinfo {author}
  {\bibfnamefont {J.}~\bibnamefont {Volz}}, \ and\ \bibinfo {author}
  {\bibfnamefont {A.}~\bibnamefont {Rauschenbeutel}},\ }\href {\doibase
  10.1103/PhysRevLett.128.073601} {\bibfield  {journal} {\bibinfo  {journal}
  {Phys. Rev. Lett.}\ }\textbf {\bibinfo {volume} {128}},\ \bibinfo {pages}
  {073601} (\bibinfo {year} {2022}{\natexlab{a}})}\BibitemShut {NoStop}%
\bibitem [{\citenamefont {Pennetta}\ \emph
  {et~al.}(2022{\natexlab{b}})\citenamefont {Pennetta}, \citenamefont
  {Lechner}, \citenamefont {Blaha}, \citenamefont {Rauschenbeutel},
  \citenamefont {Schneeweiss},\ and\ \citenamefont {Volz}}]{Pennetta2022b}%
  \BibitemOpen
  \bibfield  {author} {\bibinfo {author} {\bibfnamefont {R.}~\bibnamefont
  {Pennetta}}, \bibinfo {author} {\bibfnamefont {D.}~\bibnamefont {Lechner}},
  \bibinfo {author} {\bibfnamefont {M.}~\bibnamefont {Blaha}}, \bibinfo
  {author} {\bibfnamefont {A.}~\bibnamefont {Rauschenbeutel}}, \bibinfo
  {author} {\bibfnamefont {P.}~\bibnamefont {Schneeweiss}}, \ and\ \bibinfo
  {author} {\bibfnamefont {J.}~\bibnamefont {Volz}},\ }\href {\doibase
  10.1103/PhysRevLett.128.203601} {\bibfield  {journal} {\bibinfo  {journal}
  {Phys. Rev. Lett.}\ }\textbf {\bibinfo {volume} {128}},\ \bibinfo {pages}
  {203601} (\bibinfo {year} {2022}{\natexlab{b}})}\BibitemShut {NoStop}%
\bibitem [{\citenamefont {Kumlin}\ \emph {et~al.}(2020)\citenamefont {Kumlin},
  \citenamefont {Kleinbeck}, \citenamefont {Stiesdal}, \citenamefont {Busche},
  \citenamefont {Hofferberth},\ and\ \citenamefont {B\"uchler}}]{Kumlin2020}%
  \BibitemOpen
  \bibfield  {author} {\bibinfo {author} {\bibfnamefont {J.}~\bibnamefont
  {Kumlin}}, \bibinfo {author} {\bibfnamefont {K.}~\bibnamefont {Kleinbeck}},
  \bibinfo {author} {\bibfnamefont {N.}~\bibnamefont {Stiesdal}}, \bibinfo
  {author} {\bibfnamefont {H.}~\bibnamefont {Busche}}, \bibinfo {author}
  {\bibfnamefont {S.}~\bibnamefont {Hofferberth}}, \ and\ \bibinfo {author}
  {\bibfnamefont {H.~P.}\ \bibnamefont {B\"uchler}},\ }\href {\doibase
  10.1103/PhysRevA.102.063703} {\bibfield  {journal} {\bibinfo  {journal}
  {Phys. Rev. A}\ }\textbf {\bibinfo {volume} {102}},\ \bibinfo {pages}
  {063703} (\bibinfo {year} {2020})}\BibitemShut {NoStop}%
\bibitem [{\citenamefont {Solano}\ \emph
  {et~al.}(2017{\natexlab{a}})\citenamefont {Solano}, \citenamefont {Grover},
  \citenamefont {Hoffman}, \citenamefont {Ravets}, \citenamefont {Fatemi},
  \citenamefont {Orozco},\ and\ \citenamefont {Rolston}}]{solano17d}%
  \BibitemOpen
  \bibfield  {author} {\bibinfo {author} {\bibfnamefont {P.}~\bibnamefont
  {Solano}}, \bibinfo {author} {\bibfnamefont {J.~A.}\ \bibnamefont {Grover}},
  \bibinfo {author} {\bibfnamefont {J.~E.}\ \bibnamefont {Hoffman}}, \bibinfo
  {author} {\bibfnamefont {S.}~\bibnamefont {Ravets}}, \bibinfo {author}
  {\bibfnamefont {F.~K.}\ \bibnamefont {Fatemi}}, \bibinfo {author}
  {\bibfnamefont {L.~A.}\ \bibnamefont {Orozco}}, \ and\ \bibinfo {author}
  {\bibfnamefont {S.~L.}\ \bibnamefont {Rolston}},\ }in\ \href@noop {} {\emph
  {\bibinfo {booktitle} {Advances In Atomic, Molecular, and Optical
  Physics}}},\ Vol.~\bibinfo {volume} {66}\ (\bibinfo  {publisher} {Elsevier},\
  \bibinfo {year} {2017})\ pp.\ \bibinfo {pages} {439--505}\BibitemShut
  {NoStop}%
\bibitem [{\citenamefont {Hoffman}\ \emph {et~al.}(2014)\citenamefont
  {Hoffman}, \citenamefont {Ravets}, \citenamefont {Grover}, \citenamefont
  {Solano}, \citenamefont {Kordell}, \citenamefont {Wong-Campos}, \citenamefont
  {Orozco},\ and\ \citenamefont {Rolston}}]{hoffman14}%
  \BibitemOpen
  \bibfield  {author} {\bibinfo {author} {\bibfnamefont {J.~E.}\ \bibnamefont
  {Hoffman}}, \bibinfo {author} {\bibfnamefont {S.}~\bibnamefont {Ravets}},
  \bibinfo {author} {\bibfnamefont {J.~A.}\ \bibnamefont {Grover}}, \bibinfo
  {author} {\bibfnamefont {P.}~\bibnamefont {Solano}}, \bibinfo {author}
  {\bibfnamefont {P.~R.}\ \bibnamefont {Kordell}}, \bibinfo {author}
  {\bibfnamefont {J.~D.}\ \bibnamefont {Wong-Campos}}, \bibinfo {author}
  {\bibfnamefont {L.~A.}\ \bibnamefont {Orozco}}, \ and\ \bibinfo {author}
  {\bibfnamefont {S.~L.}\ \bibnamefont {Rolston}},\ }\href {\doibase
  10.1063/1.4879799} {\bibfield  {journal} {\bibinfo  {journal} {AIP Advances}\
  }\textbf {\bibinfo {volume} {4}},\ \bibinfo {pages} {067124} (\bibinfo {year}
  {2014})}\BibitemShut {NoStop}%
\bibitem [{\citenamefont {Vetsch}\ \emph {et~al.}(2010)\citenamefont {Vetsch},
  \citenamefont {Reitz}, \citenamefont {Sagu\'e}, \citenamefont {Schmidt},
  \citenamefont {Dawkins},\ and\ \citenamefont {Rauschenbeutel}}]{Vetsch2010}%
  \BibitemOpen
  \bibfield  {author} {\bibinfo {author} {\bibfnamefont {E.}~\bibnamefont
  {Vetsch}}, \bibinfo {author} {\bibfnamefont {D.}~\bibnamefont {Reitz}},
  \bibinfo {author} {\bibfnamefont {G.}~\bibnamefont {Sagu\'e}}, \bibinfo
  {author} {\bibfnamefont {R.}~\bibnamefont {Schmidt}}, \bibinfo {author}
  {\bibfnamefont {S.~T.}\ \bibnamefont {Dawkins}}, \ and\ \bibinfo {author}
  {\bibfnamefont {A.}~\bibnamefont {Rauschenbeutel}},\ }\href {\doibase
  10.1103/PhysRevLett.104.203603} {\bibfield  {journal} {\bibinfo  {journal}
  {Phys. Rev. Lett.}\ }\textbf {\bibinfo {volume} {104}},\ \bibinfo {pages}
  {203603} (\bibinfo {year} {2010})}\BibitemShut {NoStop}%
\bibitem [{\citenamefont {Goban}\ \emph {et~al.}(2012)\citenamefont {Goban},
  \citenamefont {Choi}, \citenamefont {Alton}, \citenamefont {Ding},
  \citenamefont {Lacro\^ute}, \citenamefont {Pototschnig}, \citenamefont
  {Thiele}, \citenamefont {Stern},\ and\ \citenamefont {Kimble}}]{Goban2012}%
  \BibitemOpen
  \bibfield  {author} {\bibinfo {author} {\bibfnamefont {A.}~\bibnamefont
  {Goban}}, \bibinfo {author} {\bibfnamefont {K.~S.}\ \bibnamefont {Choi}},
  \bibinfo {author} {\bibfnamefont {D.~J.}\ \bibnamefont {Alton}}, \bibinfo
  {author} {\bibfnamefont {D.}~\bibnamefont {Ding}}, \bibinfo {author}
  {\bibfnamefont {C.}~\bibnamefont {Lacro\^ute}}, \bibinfo {author}
  {\bibfnamefont {M.}~\bibnamefont {Pototschnig}}, \bibinfo {author}
  {\bibfnamefont {T.}~\bibnamefont {Thiele}}, \bibinfo {author} {\bibfnamefont
  {N.~P.}\ \bibnamefont {Stern}}, \ and\ \bibinfo {author} {\bibfnamefont
  {H.~J.}\ \bibnamefont {Kimble}},\ }\href {\doibase
  10.1103/PhysRevLett.109.033603} {\bibfield  {journal} {\bibinfo  {journal}
  {Phys. Rev. Lett.}\ }\textbf {\bibinfo {volume} {109}},\ \bibinfo {pages}
  {033603} (\bibinfo {year} {2012})}\BibitemShut {NoStop}%
\bibitem [{\citenamefont {Ding}\ \emph {et~al.}(2012)\citenamefont {Ding},
  \citenamefont {Goban}, \citenamefont {Choi},\ and\ \citenamefont
  {Kimble}}]{Gobanerrata2012}%
  \BibitemOpen
  \bibfield  {author} {\bibinfo {author} {\bibfnamefont {D.}~\bibnamefont
  {Ding}}, \bibinfo {author} {\bibfnamefont {A.}~\bibnamefont {Goban}},
  \bibinfo {author} {\bibfnamefont {K.~S.}\ \bibnamefont {Choi}}, \ and\
  \bibinfo {author} {\bibfnamefont {H.~J.}\ \bibnamefont {Kimble}},\ }\href
  {https://arxiv.org/abs/1212.4941} {\bibfield  {journal} {\bibinfo  {journal}
  {arXiv}\ }\textbf {\bibinfo {volume} {{ }}},\ \bibinfo {pages} {1212.4941}
  (\bibinfo {year} {2012})}\BibitemShut {NoStop}%
\bibitem [{\citenamefont {Kaluzny}\ \emph {et~al.}(1983)\citenamefont
  {Kaluzny}, \citenamefont {Goy}, \citenamefont {Gross}, \citenamefont
  {Raimond},\ and\ \citenamefont {Haroche}}]{Kaluzny1983}%
  \BibitemOpen
  \bibfield  {author} {\bibinfo {author} {\bibfnamefont {Y.}~\bibnamefont
  {Kaluzny}}, \bibinfo {author} {\bibfnamefont {P.}~\bibnamefont {Goy}},
  \bibinfo {author} {\bibfnamefont {M.}~\bibnamefont {Gross}}, \bibinfo
  {author} {\bibfnamefont {J.~M.}\ \bibnamefont {Raimond}}, \ and\ \bibinfo
  {author} {\bibfnamefont {S.}~\bibnamefont {Haroche}},\ }\href {\doibase
  10.1103/PhysRevLett.51.1175} {\bibfield  {journal} {\bibinfo  {journal}
  {Phys. Rev. Lett.}\ }\textbf {\bibinfo {volume} {51}},\ \bibinfo {pages}
  {1175} (\bibinfo {year} {1983})}\BibitemShut {NoStop}%
\bibitem [{\citenamefont {Mielke}\ \emph {et~al.}(1997)\citenamefont {Mielke},
  \citenamefont {Foster}, \citenamefont {Gripp},\ and\ \citenamefont
  {Orozco}}]{Mielke1997}%
  \BibitemOpen
  \bibfield  {author} {\bibinfo {author} {\bibfnamefont {S.~L.}\ \bibnamefont
  {Mielke}}, \bibinfo {author} {\bibfnamefont {G.~T.}\ \bibnamefont {Foster}},
  \bibinfo {author} {\bibfnamefont {J.}~\bibnamefont {Gripp}}, \ and\ \bibinfo
  {author} {\bibfnamefont {L.~A.}\ \bibnamefont {Orozco}},\ }\href {\doibase
  10.1364/OL.22.000325} {\bibfield  {journal} {\bibinfo  {journal} {Opt.
  Lett.}\ }\textbf {\bibinfo {volume} {22}},\ \bibinfo {pages} {325} (\bibinfo
  {year} {1997})}\BibitemShut {NoStop}%
\bibitem [{\citenamefont {Carmichael}\ \emph {et~al.}(1991)\citenamefont
  {Carmichael}, \citenamefont {Brecha},\ and\ \citenamefont
  {Rice}}]{carmichael91}%
  \BibitemOpen
  \bibfield  {author} {\bibinfo {author} {\bibfnamefont {H.}~\bibnamefont
  {Carmichael}}, \bibinfo {author} {\bibfnamefont {R.}~\bibnamefont {Brecha}},
  \ and\ \bibinfo {author} {\bibfnamefont {P.}~\bibnamefont {Rice}},\ }\href
  {\doibase https://doi.org/10.1016/0030-4018(91)90194-I} {\bibfield  {journal}
  {\bibinfo  {journal} {Optics Communications}\ }\textbf {\bibinfo {volume}
  {82}},\ \bibinfo {pages} {73} (\bibinfo {year} {1991})}\BibitemShut {NoStop}%
\bibitem [{\citenamefont {Asenjo-Garcia}\ \emph {et~al.}(2017)\citenamefont
  {Asenjo-Garcia}, \citenamefont {Hood}, \citenamefont {Chang},\ and\
  \citenamefont {Kimble}}]{Asenjo2017}%
  \BibitemOpen
  \bibfield  {author} {\bibinfo {author} {\bibfnamefont {A.}~\bibnamefont
  {Asenjo-Garcia}}, \bibinfo {author} {\bibfnamefont {J.~D.}\ \bibnamefont
  {Hood}}, \bibinfo {author} {\bibfnamefont {D.~E.}\ \bibnamefont {Chang}}, \
  and\ \bibinfo {author} {\bibfnamefont {H.~J.}\ \bibnamefont {Kimble}},\
  }\href {\doibase 10.1103/PhysRevA.95.033818} {\bibfield  {journal} {\bibinfo
  {journal} {Phys. Rev. A}\ }\textbf {\bibinfo {volume} {95}},\ \bibinfo
  {pages} {033818} (\bibinfo {year} {2017})}\BibitemShut {NoStop}%
\bibitem [{\citenamefont {Solano}\ \emph
  {et~al.}(2017{\natexlab{b}})\citenamefont {Solano}, \citenamefont
  {Barberis-Blostein}, \citenamefont {Fatemi}, \citenamefont {Orozco},\ and\
  \citenamefont {Rolston}}]{solano17a}%
  \BibitemOpen
  \bibfield  {author} {\bibinfo {author} {\bibfnamefont {P.}~\bibnamefont
  {Solano}}, \bibinfo {author} {\bibfnamefont {P.}~\bibnamefont
  {Barberis-Blostein}}, \bibinfo {author} {\bibfnamefont {F.~K.}\ \bibnamefont
  {Fatemi}}, \bibinfo {author} {\bibfnamefont {L.~A.}\ \bibnamefont {Orozco}},
  \ and\ \bibinfo {author} {\bibfnamefont {S.~L.}\ \bibnamefont {Rolston}},\
  }\href {\doibase 10.1038/s41467-017-01994-3} {\bibfield  {journal} {\bibinfo
  {journal} {Nature Communications}\ }\textbf {\bibinfo {volume} {8}},\
  \bibinfo {pages} {1857} (\bibinfo {year} {2017}{\natexlab{b}})}\BibitemShut
  {NoStop}%
\bibitem [{SI()}]{SI}%
  \BibitemOpen
  \href@noop {} {}\bibinfo {note} {See Supplementary Information for further
  details on the theory model, transmitted intensity in terms of collective
  modes and in the continuous limit, absorption sprectrum and data
  processing}\BibitemShut {NoStop}%
\bibitem [{\citenamefont {Deutsch}\ \emph {et~al.}(1995)\citenamefont
  {Deutsch}, \citenamefont {Spreeuw}, \citenamefont {Rolston},\ and\
  \citenamefont {Phillips}}]{Deutsch1995}%
  \BibitemOpen
  \bibfield  {author} {\bibinfo {author} {\bibfnamefont {I.~H.}\ \bibnamefont
  {Deutsch}}, \bibinfo {author} {\bibfnamefont {R.~J.~C.}\ \bibnamefont
  {Spreeuw}}, \bibinfo {author} {\bibfnamefont {S.~L.}\ \bibnamefont
  {Rolston}}, \ and\ \bibinfo {author} {\bibfnamefont {W.~D.}\ \bibnamefont
  {Phillips}},\ }\href {\doibase 10.1103/PhysRevA.52.1394} {\bibfield
  {journal} {\bibinfo  {journal} {Phys. Rev. A}\ }\textbf {\bibinfo {volume}
  {52}},\ \bibinfo {pages} {1394} (\bibinfo {year} {1995})}\BibitemShut
  {NoStop}%
\bibitem [{\citenamefont {Corzo}\ \emph {et~al.}(2016)\citenamefont {Corzo},
  \citenamefont {Gouraud}, \citenamefont {Chandra}, \citenamefont {Goban},
  \citenamefont {Sheremet}, \citenamefont {Kupriyanov},\ and\ \citenamefont
  {Laurat}}]{Corzo2016}%
  \BibitemOpen
  \bibfield  {author} {\bibinfo {author} {\bibfnamefont {N.~V.}\ \bibnamefont
  {Corzo}}, \bibinfo {author} {\bibfnamefont {B.}~\bibnamefont {Gouraud}},
  \bibinfo {author} {\bibfnamefont {A.}~\bibnamefont {Chandra}}, \bibinfo
  {author} {\bibfnamefont {A.}~\bibnamefont {Goban}}, \bibinfo {author}
  {\bibfnamefont {A.~S.}\ \bibnamefont {Sheremet}}, \bibinfo {author}
  {\bibfnamefont {D.}~\bibnamefont {Kupriyanov}}, \ and\ \bibinfo {author}
  {\bibfnamefont {J.}~\bibnamefont {Laurat}},\ }\href {\doibase
  10.1103/PhysRevLett.117.133603} {\bibfield  {journal} {\bibinfo  {journal}
  {Phys. Rev. Lett.}\ }\textbf {\bibinfo {volume} {117}},\ \bibinfo {pages}
  {133603} (\bibinfo {year} {2016})}\BibitemShut {NoStop}%
\bibitem [{\citenamefont {S\o{}rensen}\ \emph {et~al.}(2016)\citenamefont
  {S\o{}rensen}, \citenamefont {B\'eguin}, \citenamefont {Kluge}, \citenamefont
  {Iakoupov}, \citenamefont {S\o{}rensen}, \citenamefont {M\"uller},
  \citenamefont {Polzik},\ and\ \citenamefont {Appel}}]{Sorensen2016}%
  \BibitemOpen
  \bibfield  {author} {\bibinfo {author} {\bibfnamefont {H.~L.}\ \bibnamefont
  {S\o{}rensen}}, \bibinfo {author} {\bibfnamefont {J.-B.}\ \bibnamefont
  {B\'eguin}}, \bibinfo {author} {\bibfnamefont {K.~W.}\ \bibnamefont {Kluge}},
  \bibinfo {author} {\bibfnamefont {I.}~\bibnamefont {Iakoupov}}, \bibinfo
  {author} {\bibfnamefont {A.~S.}\ \bibnamefont {S\o{}rensen}}, \bibinfo
  {author} {\bibfnamefont {J.~H.}\ \bibnamefont {M\"uller}}, \bibinfo {author}
  {\bibfnamefont {E.~S.}\ \bibnamefont {Polzik}}, \ and\ \bibinfo {author}
  {\bibfnamefont {J.}~\bibnamefont {Appel}},\ }\href {\doibase
  10.1103/PhysRevLett.117.133604} {\bibfield  {journal} {\bibinfo  {journal}
  {Phys. Rev. Lett.}\ }\textbf {\bibinfo {volume} {117}},\ \bibinfo {pages}
  {133604} (\bibinfo {year} {2016})}\BibitemShut {NoStop}%
\bibitem [{\citenamefont {Patterson}\ \emph {et~al.}(2018)\citenamefont
  {Patterson}, \citenamefont {Solano}, \citenamefont {Julienne}, \citenamefont
  {Orozco},\ and\ \citenamefont {Rolston}}]{Patterson2018}%
  \BibitemOpen
  \bibfield  {author} {\bibinfo {author} {\bibfnamefont {B.~D.}\ \bibnamefont
  {Patterson}}, \bibinfo {author} {\bibfnamefont {P.}~\bibnamefont {Solano}},
  \bibinfo {author} {\bibfnamefont {P.~S.}\ \bibnamefont {Julienne}}, \bibinfo
  {author} {\bibfnamefont {L.~A.}\ \bibnamefont {Orozco}}, \ and\ \bibinfo
  {author} {\bibfnamefont {S.~L.}\ \bibnamefont {Rolston}},\ }\href {\doibase
  10.1103/PhysRevA.97.032509} {\bibfield  {journal} {\bibinfo  {journal} {Phys.
  Rev. A}\ }\textbf {\bibinfo {volume} {97}},\ \bibinfo {pages} {032509}
  (\bibinfo {year} {2018})}\BibitemShut {NoStop}%
\bibitem [{\citenamefont {Su}\ \emph {et~al.}(2022)\citenamefont {Su},
  \citenamefont {Solano}, \citenamefont {Wack}, \citenamefont {Orozco},\ and\
  \citenamefont {Zhao}}]{Su2022}%
  \BibitemOpen
  \bibfield  {author} {\bibinfo {author} {\bibfnamefont {D.}~\bibnamefont
  {Su}}, \bibinfo {author} {\bibfnamefont {P.}~\bibnamefont {Solano}}, \bibinfo
  {author} {\bibfnamefont {J.~D.}\ \bibnamefont {Wack}}, \bibinfo {author}
  {\bibfnamefont {L.~A.}\ \bibnamefont {Orozco}}, \ and\ \bibinfo {author}
  {\bibfnamefont {Y.}~\bibnamefont {Zhao}},\ }\href {\doibase
  10.1364/PRJ.440991} {\bibfield  {journal} {\bibinfo  {journal} {Photon.
  Res.}\ }\textbf {\bibinfo {volume} {10}},\ \bibinfo {pages} {601} (\bibinfo
  {year} {2022})}\BibitemShut {NoStop}%
\bibitem [{\citenamefont {Wei}\ \emph {et~al.}(2009)\citenamefont {Wei},
  \citenamefont {Chen}, \citenamefont {Loy}, \citenamefont {Wong},\ and\
  \citenamefont {Du}}]{Wei2009}%
  \BibitemOpen
  \bibfield  {author} {\bibinfo {author} {\bibfnamefont {D.}~\bibnamefont
  {Wei}}, \bibinfo {author} {\bibfnamefont {J.~F.}\ \bibnamefont {Chen}},
  \bibinfo {author} {\bibfnamefont {M.~M.~T.}\ \bibnamefont {Loy}}, \bibinfo
  {author} {\bibfnamefont {G.~K.~L.}\ \bibnamefont {Wong}}, \ and\ \bibinfo
  {author} {\bibfnamefont {S.}~\bibnamefont {Du}},\ }\href {\doibase
  10.1103/PhysRevLett.103.093602} {\bibfield  {journal} {\bibinfo  {journal}
  {Phys. Rev. Lett.}\ }\textbf {\bibinfo {volume} {103}},\ \bibinfo {pages}
  {093602} (\bibinfo {year} {2009})}\BibitemShut {NoStop}%
\end{thebibliography}%
\clearpage
\onecolumngrid
\section{SUPPLEMENTARY INFORMATION}

\subsection{1. Theory}
\subsubsection{1.1. Many-atom theory}

An ensemble of $N$ two-level atoms coupled to a waveguide and driven by a weak pulse with an arbitrary temporal shape $\Omega(t)$ is described by the effective Hamiltonian $\mathcal{H}=\mathcal{H}_\text{1D}+\mathcal{H}'+\mathcal{H}_\text{drive}$, where 
\begin{subequations}
\begin{gather}
\mathcal{H}_\text{1D}=-i \frac{\hbar\Gamma_\text{1D}}{2}\sum_{i,j=1}^N e^{ik d |i-j|} \hat{\sigma}^i_{eg}\hat{\sigma}^j_{ge}, \\
\mathcal{H}'=-i\frac{\hbar\Gamma'}{2}\sum_{i=1}^N\hat{\sigma}^i_{ee},\\
\mathcal{H}_\text{drive}=-\hbar\Delta\sum_{i=1}^N \hat{\sigma}^i_{ee}-\hbar\Omega(t)\sum_{i=1}^N \left(e^{i k z_i}\hat{\sigma}^i_{eg} +\text{H. c.}\right).
\end{gather}
\label{eq:ham}
\end{subequations}
Here, $k$, $\Delta$ and $\{z_i\}$ are respectively the light wavevector, the detuning between the central frequency of the drive and the atomic resonance, and the set of atomic positions along the waveguide. The expectation value of the atomic coherences $\langle\hat{\sigma}^n_{ge}\rangle\equiv\sigma_{ge}^n $ evolves according to
\begin{equation}
\dot{\sigma}_{ge}^n=i\bigg(\Delta+i\frac{\Gamma'}{2}\bigg)\sigma_{ge}^n+i \Omega(t)e^{i k z_n}-\frac{\Gamma_\text{1D}}{2}\sum_{m=1}^N e^{ik|z_n-z_m|}\sigma_{ge}^m\label{sims}.
\end{equation}

The  expectation value for (the positive-frequency component of) the electric field at a point to the left of the ensemble is a sum of the input field and the field generated by the atoms,  
\begin{equation}
   E^+(z, t)=\Omega(t) e^{ik z}+i\frac{\Gamma_\text{1D}}{2}\sum_{i=1}^N e^{ik |z-z_i|}\sigma_{ge}^i(t).
   \label{sims_field}
\end{equation}

The results presented in Fig. 2\textbf{(b)} are calculated by numerically solving Eq.~\eqref{sims} using the amplitude of the experimental pulse as $\Omega(t)$, for 40 disordered atoms with $OD=11.6$  and $\Gamma_{1D}/\Gamma'=OD/2N$.  As discussed in the main text, the output field is determined solely by $N\Gamma_{1D}$ in the large atom limit, so the specific number of atoms and its configuration are irrelevant as long as $N\gg 1$. Noise has been added to the final result to make the experimental signal and the simulation agree at long times (due to the logarithmic scale, this is a small correction at short times).
 
 \subsubsection{1.2. Transmitted intensity in terms of collective modes}
 
 The system of equations in Eq.~\eqref{sims} is simpler if we  transform to the basis that diagonalizes the atom-atom interactions. In this representation, Eq.~\eqref{sims} becomes
 \begin{equation}
\dot{\tilde{\sigma}}_{ge}^{\xi}+(\Gamma'/2+i\lambda_\xi-i\Delta)\tilde{\sigma}_{ge}^{\xi}=i\tilde{\Omega}^\xi,
\label{modes}
\end{equation}
where $\{\lambda_\xi\}$ are the eigenvalues of the Hamiltonian of Eq.~\eqref{eq:ham}(a) in the single excitation sector. The real ($J_\xi=\text{Re}(\lambda_\xi)$) and imaginary ($\Gamma_\xi=-2\text{Im}(\lambda_\xi)$) parts of these eigenvalues encode the frequency shift and collective decay rate of each of the modes. In the above equation, $\tilde{\sigma}_{ge}^{\xi}=\sum_{i=1}^N v_\xi^i\sigma_{ge}^i$, and $\tilde{\Omega}^{\xi}=\sum_{i=1}^N v_\xi^i\Omega_{ge}^i$ are linear combinations of the coherences and the field at the atomic positions, projected onto the eigenvectors $\textbf{v}_\xi$. The dynamics is thus that of $N$ independent dipoles.

If atoms are in the mirror configuration, there is  a single bright mode with eigenvalue $\lambda_\text{bright}=-i N\Gamma_\text{1D}/2$. The above equation becomes 
\begin{equation}
\dot{p}+(\Gamma'/2+OD\Gamma'/4 -i\Delta)p=i\Omega_\text{bright},
\end{equation}
where we have defined $p\equiv\tilde{\sigma}_{eg}^\text{bright}$. On resonance (with $\Delta=0$), this equation is equivalent to Eq.~(1) in the main text.

All collective modes contribute to the output field if atoms are not in the mirror configuration. To calculate the output field, we use the transmission coefficient~\cite{Asenjo2017}  
\begin{equation}
    T(\omega)=1-\frac{i\Gamma_{1D}}{2}\sum_{\xi=1}^N \frac{\eta_\xi}{\omega-\omega_0+i\Gamma'/2-\lambda_\xi},
    \label{TN}
\end{equation}
which is obtained by solving Eq.~\eqref{sims} in the steady state and plugging the solution for the coherences in Eq.~\eqref{sims_field}. Here, $\{\eta_\xi\}$ are coefficients given in terms of the eigenvectors $\{\textbf{v}_\xi\}$ as
\begin{equation}
    \eta_\xi=\sum_{n=1}^N\sum_{m=1}^N v_{\xi,n}v_{\xi,m}e^{-ikd(n-m)}.
\end{equation}

We model the input pulse as a Gaussian with central frequency $\omega_p$ and standard deviation $\sigma={\rm{FWHM}}/{2\sqrt{2\ln 2}}$. Plugging Eq.~\eqref{TN} into Eq.~\eqref{FT of Transmission} yields
\begin{equation}
    \frac{I_d(t)}{I_0}=\bigg|E_0(t)e^{-i\Delta t}-\frac{\Gamma_{1D}}{4}\sum_{\xi=1}^N \eta_\xi e^{-(\Gamma'+\Gamma_{\xi})t/2}e^{-i J_\xi t}e^{-\frac{1}{2}(\Delta+i\Gamma'/2-\lambda_\xi)^2\sigma^2}\text{erfc}\left(\frac{-i(\Delta+i\Gamma'/2-\lambda_\xi)\sigma^2-t}{\sqrt{2}\sigma}\right)\bigg|^2,
\end{equation}
where $\Delta\equiv\omega_p-\omega_0$. The coefficients $a_\xi$ and $b_\xi$ of Eq.~\eqref{eq:IntensityNmodes} thus read
\begin{subequations}
\begin{gather}
a_\xi=\eta_\xi e^{-\frac{1}{2}(\Delta+i\Gamma'/2-\lambda_\xi)^2\sigma^2},\\
b_\xi=-i(\Delta+i\Gamma'/2-\lambda_\xi)\sigma^2.
\end{gather}
\label{eq:inmodes}
\end{subequations}
The coefficients $\eta_\xi$ weight the contributions of different modes. The most superradiant modes have a larger coefficient and contribute more to the final signal.

For late times we can approximate $\text{erfc}(\cdot)\sim 2$ for all the terms. Furthermore, we neglect the input field to determine the slope of the second peak and approximate the scattered field by the term corresponding to the first few most superradiant modes:
\begin{equation}
    \frac{I_d(t)}{I_0}\propto\bigg|\sum_{\xi=1}^{\xi_\text{cut}} \tilde{\eta}_\xi e^{-(\Gamma'+\Gamma_{\xi})t/2}e^{-i J_\xi t}\bigg|^2.
\end{equation}
In the above equation, $\xi_\text{cut}\in\mathbb{N}$ is a mode cut-off, and $\tilde{\eta}_\xi$ is a prefactor absorbing all the constants. The first few superradiant modes have an scaling $\Gamma_{\xi}\sim N\Gamma_\text{1D}$. We approximate $\Gamma_\xi\sim N\Gamma_\text{1D}$ and write
\begin{equation}
    \frac{I_d(t)}{I_0}\propto e^{-(\Gamma'+N\Gamma_\text{1D})t}\bigg|\sum_{\xi=1}^{\xi_\text{cut}} \tilde{\eta}_\xi e^{-\Delta_{\xi} t/2}e^{-i J_\xi t}\bigg|^2.
\end{equation}
Here $\Delta_{\xi}$ is a correction that arises from the approximation of $\Gamma_{\xi}$ as $N\Gamma_\text{1D}$. Overall $I_d(t)\propto I_0 e^{-(\Gamma'+N\Gamma_{1D})t}F(t)$, with $F(t)$ a function whose evolution timescale is much larger than $(\Gamma'+N\Gamma_\text{1D})^{-1}$. The effective decay rate at the second maximum is then approximately $\Gamma'+N\Gamma_\text{1D}=\Gamma'(1+\frac{OD}{2})$. Note that this approximation becomes exact only for atoms in the mirror configuration.
 
 \subsubsection{1.3. Transmitted intensity in the continuous limit}

Here we provide the derivation of Eq.~(\ref{eq:FinalIntensity}) in the main text. Since $N\gg 1$, $\mathcal{T}_N(\omega)$ is well approximated by Eq. \eqref{eq:transmission} of the main text \cite{cardenas2023}. Defining $z\equiv\omega-\omega_p$, we can write the output intensity from Eq.~(\ref{FT of Transmission}) as
\begin{equation}
\frac{I_d(t)}{I_0}=\left|\frac{1}{2\pi}\int_{-\infty}^\infty  \text{exp}\left(\frac{-i N\Gamma_{1D}}{2}\frac{1}{z+\Delta+i\Gamma'/2}\right) \text{exp}\left(-\frac{1}{2}\sigma^2 z^2\right) \text{exp}\left(-i z t\right)dz\right|^2.
\label{eq:presc2}
\end{equation}

Via the Bessel generating function,
\begin{equation}
\begin{aligned}
&\text{exp}\left(\frac{-i N\Gamma_{1D}}{2}\frac{1}{z+\Delta+i\Gamma'/2}-i z t\right)=e^{\left(i\Delta-\frac{\Gamma'}{2}\right) t}\sum_{m=-\infty}^\infty \left(-i\left(z+\Delta+i \frac{\Gamma'}{2}\right)\sqrt{\frac{2t}{N\Gamma_{1D}}}\right)^m J_m\left(\sqrt{2N\Gamma_{1D} t}\right),
\end{aligned}
\end{equation}
the output intensity can be expressed as the expansion given in Eq. (\ref{eq:FinalIntensity}). The expansion coefficients take the form
\begin{equation}
    A_m=\frac{(-2i)^m}{2\pi}\int_{-\infty}^\infty dz \left(z+\Delta+i \frac{\Gamma'}{2}\right)^m e^{-\frac{1}{2}\sigma^2 z^2}.
    \label{coefs}
\end{equation}

For $m>0$, $A_m$ can be written in terms of a confluent hypergeometric function of the second kind
\begin{equation}
    A_m= \frac{(2)^{m-1}}{\sqrt{\pi}} \left(\frac{2}{\sigma^2}\right)^{\frac{m+1}{2}}U\left(-\frac{1}{2}m;\frac{1}{2};-\frac{\sigma^2(\Delta+i \frac{\Gamma'}{2})^2}{2}\right).
        \label{coefs1}
\end{equation}
For $m<0$,
\begin{equation}
    A_m= \frac{(-2i)^m}{2\pi}\frac{(-1)^{|m|-1}}{(|m|-1)!}\partial_\alpha^{|m|-1} F(\alpha,\sigma)\big|_{\alpha=\Delta+i\frac{\Gamma'}{2}},
    \label{coefs2}
\end{equation}
with
\begin{equation}
F(\alpha,\sigma)=-i\pi e^{-\frac{1}{2}\sigma^2\alpha^2}\left(\text{erf}\left(i\frac{\alpha\sigma}{\sqrt{2}}\right)+1\right).
\end{equation}

\subsection{2. Absorption spectrum}

The optical dipole trap produces position-dependent light-shifts. The transmission coefficient, given by the Beer-Lambert law, of the actual atomic sample departs from the exponential of a Lorentzian distribution, as in Eq. (3) of the main text. As a result, the absorption spectrum (i.e., the transmittance) can often look broadened and even asymmetric. 

The details of the position distribution of the atoms and position-dependent light-shifts are not directly measurable and difficult to estimate. We have the measured transmittance of the intensity, but we need that of the field with its real and imaginary components. We use a phenomenological approach to describe the absorption coefficient of the total atomic sample, $\mathcal{T}(\omega)$, by considering it as an ensemble of smaller samples with optical density $OD_i$ frequency-shifted by $\omega_i$ with transmission coefficient $\mathcal{T}_i(\omega,\omega_i,OD_i)$. This is $\mathcal{T}(\omega)=\prod_i \mathcal{T}_i(\omega,\omega_i,OD_i)$. All the shifts are randomly distributed, as we assume that the position distribution is random within the traps but bounded to a maximum shift on one side of the spectrum, a process characterized by log-normal distributions. 

\begin{figure}[h]
\begin{center}
\includegraphics[width=0.5\textwidth]{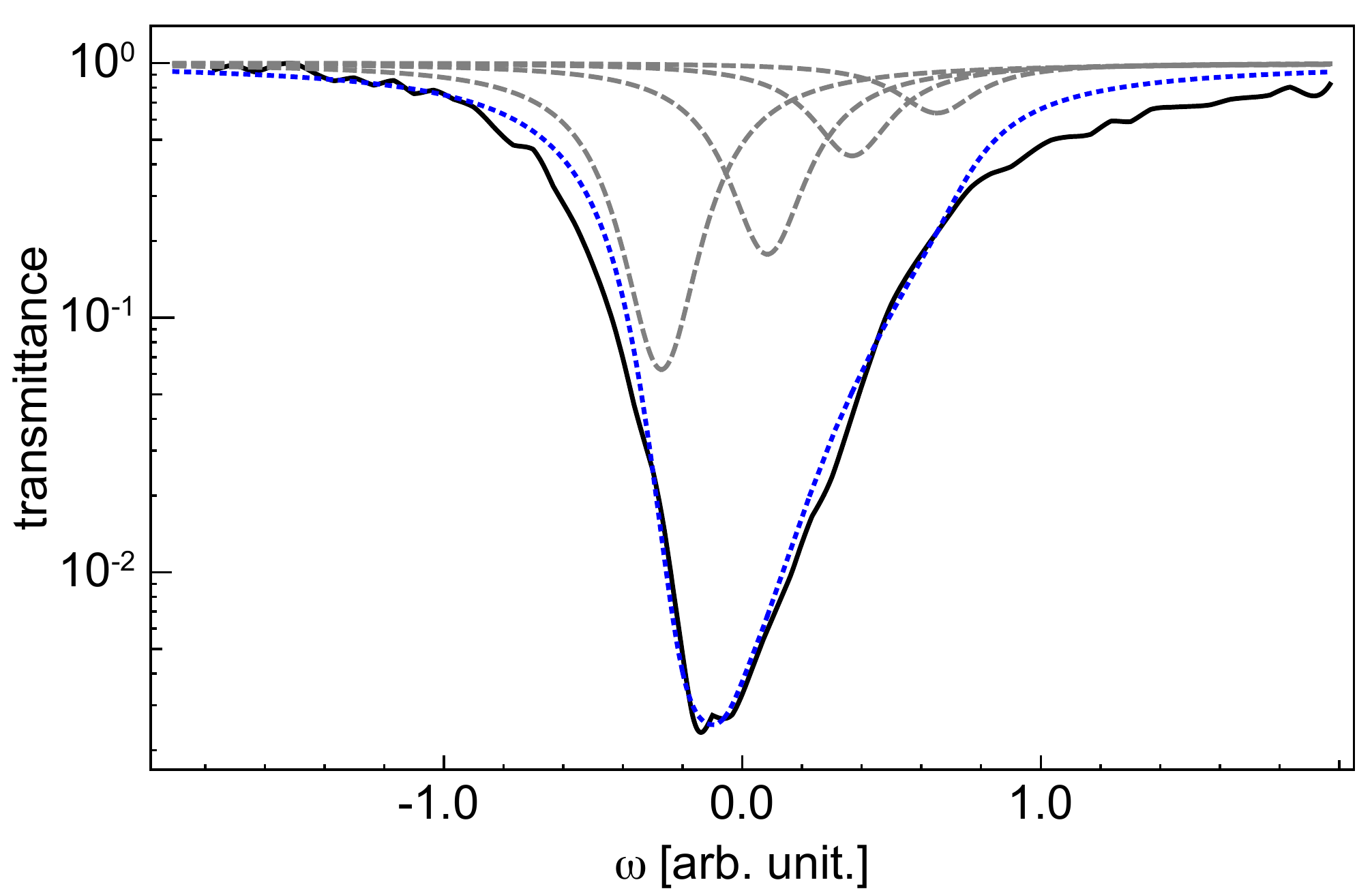}
\caption{Comparison of the measured transmittance (solid black in logarithmic scale) with the modeled one (dotted blue). The model consist of a sum of frequency shifted Lorentzian spectra (dashed gray) log-normally distributed.}
\label{fig:SM}
\end{center}
\end{figure}

Figure \ref{fig:SM} presents an example of a few light-shifted transmission spectra sampled over a log-normal distribution that produce an asymmetrically broadened transmission spectrum close to the measured one. In this case, the modeled transmission spectrum was adjusted to overlap with the measured spectrum, while in the main text, it was adjusted to predict the transmitted pulse better. A different number of sub-samples, shifts, and shapes of the log-normal distribution produces similar pulse intensity outputs as long as the transmittance roughly overlaps with the measured one. These results suggest that an approximate model is enough to reproduce the transmitted pulse shape, even when the actual absorption coefficient is unknown.

\subsection{3. Data processing}
The pulse transmission raw data obtained by photon counting on the TSPCM has a background with no atoms of $1.5\times 10^{-3}$ of the peak height. The equivalent background with atoms is closer to $0.5 \times 10^{-3}$. We have adjusted the background (adding some ten counts to every bin) in Fig.~\ref{Fig:3first-zero-slope} to make the traces converge to the same average value. The theoretical predictions have also had the background adjusted. The results of Fig. ~\ref{Fig:3first-zero-slope} are independent of this correction.   
\end{document}